\documentclass[twocolumn,superscriptaddress]{revtex4-1}
\usepackage[final]{graphicx}
\usepackage{amsmath}
\usepackage{hyperref}
\hypersetup{colorlinks=true, citecolor=blue, urlcolor=blue, linkcolor=blue}
\usepackage{color}	
\newcommand{\BS}{Bi$_2$Se$_3$}

\begin{document}

\title{Band structure of Bi surfaces formed on \BS{} upon exposure to air}

\author{Alexandre Gauthier}
\affiliation{Stanford Institute for Materials and Energy Sciences, SLAC National Accelerator Laboratory, Menlo Park, California 94025, USA}
\affiliation{Geballe Laboratory for Advanced Materials, Departments of Applied Physics and Physics, Stanford University, Stanford, California 94305, USA}

\author{Jonathan A. Sobota}
\affiliation{Stanford Institute for Materials and Energy Sciences, SLAC National Accelerator Laboratory, Menlo Park, California 94025, USA}

\author{Nicolas Gauthier}
\affiliation{Stanford Institute for Materials and Energy Sciences, SLAC National Accelerator Laboratory, Menlo Park, California 94025, USA}
\affiliation{Geballe Laboratory for Advanced Materials, Departments of Applied Physics and Physics, Stanford University, Stanford, California 94305, USA}

\author{Costel R. Rotundu}
\affiliation{Stanford Institute for Materials and Energy Sciences, SLAC National Accelerator Laboratory, Menlo Park, California 94025, USA}

\author{Zhi-Xun Shen}
\email{zxshen@stanford.edu}
\affiliation{Stanford Institute for Materials and Energy Sciences, SLAC National Accelerator Laboratory, Menlo Park, California 94025, USA}
\affiliation{Geballe Laboratory for Advanced Materials, Departments of Applied Physics and Physics, Stanford University, Stanford, California 94305, USA}

\author{Patrick S. Kirchmann}
\email{kirchman@stanford.edu}
\affiliation{Stanford Institute for Materials and Energy Sciences, SLAC National Accelerator Laboratory, Menlo Park, California 94025, USA}

\begin{abstract}
    \BS{} has been the focus of intense interest over the past decade due to its topological properties. Bi surfaces are known to form on \BS{} upon exposure to atmosphere, but their electronic structure has not been investigated. We report band structure measurements of such Bi surfaces using angle-resolved photoemission spectroscopy. Measured spectra can be well explained by the band structure of a single bilayer of Bi on \BS{}, and show that Bi surfaces consistently dominate the photoemission signal for air exposure times of at least 1 hour. These results demonstrate that atmospheric effects should be taken into consideration when identifying two-dimensional transport channels, and when designing surface-sensitive measurements of \BS{}, ideally limiting air exposure to no more than a few minutes.
\end{abstract}

\date{\today}
\maketitle 

\section{Introduction}

The \BS/Bi$_2$Te$_3$ family of materials has attracted significant interest for its topological properties \cite{Fu07,Chen09,Xia09,Zhang09}. These properties have been characterized using techniques including transport \cite{Kong11,Analytis10,Cao12,Barreto14}, photoemission \cite{Chen09,Xia09,Analytis10,Sobota14}, and optical probes \cite{Laforge10,Hsieh11,Kumar11,Qi10,Luo19}. Some of these measurements are performed in atmospheric conditions, which can have an effect on the results of the experiments compared to measurements performed under vacuum \cite{Analytis10,Qi10}. It is important to consider atmospheric effects which may modify the surface properties of these topological materials, particularly when identifying two-dimensional conduction channels or when performing surface-sensitive measurements. The surface electronic structure may also impact the properties of contacts for transport measurements and device applications.

\begin{figure*}
\includegraphics[width=\textwidth]{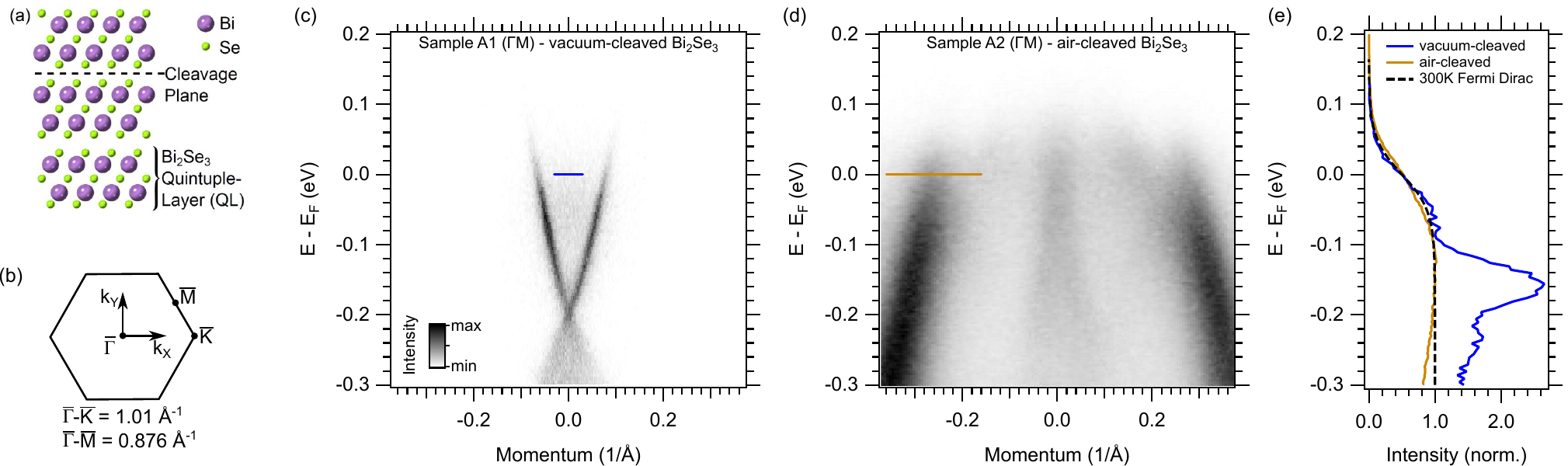}
\caption{(a) Crystal structure of \BS. (b) (111) surface Brilouin zone \cite{Nakajima63}. (c) Typical band structure along $\Gamma$-M direction of \BS{} cleaved in vacuum as measured by ARPES. (d) Band structure of \BS{}  after 1 hour of ambient air exposure. (e) Energy distribution curves for vacuum- and air-cleaved \BS{} extracted from (c) and (d). The integration range is denoted by colored bars in (c) and (d). Electronic states $E_F$ follow a Fermi-Dirac distribution for $300$~K.} 
\label{fig:arpes}
\end{figure*}

The crystal structure of \BS{} consists of layers of Se and Bi stacked in repeating Se-Bi-Se-Bi-Se quintuple layers (QLs) \cite{Nakajima63}, as depicted in Figure~\ref{fig:arpes}(a). This layered structure enables mechanical cleavage to expose fresh surfaces, both in air and in vacuum. Adjacent QLs are connected by weak van der Waals bonds between Se layers; samples cleave along these planes, resulting in Se terminated surfaces \cite{D3R13}.

Exposing \BS{} surfaces to air leads to the development of a Bi surface termination over a portion of the sample. X-ray photoelectron spectroscopy (XPS) measurements observed the surface termination of \BS{} changing from Se to Bi after air exposure of under 5 minutes \cite{Edmonds14,Hewitt14}. However, Bi surfaces were not found on every sample exposed to air \cite{Hewitt14}; this is likely due to the short air exposure time and a stochastic nature of the surface changes. 

Atomic force microscopy (AFM) measurements observed islands of height $3.2\pm2$~{\AA} that form on \BS{} upon exposure to air of at least 30 minutes \cite{Green16}. These islands can be attributed to patches of Bi forming on the surface, as this height is close to the $4$~{\AA} height of a single Bi bilayer on top of \BS{} \cite{Gibson13,Gonalves18} or sister material Bi$_2$Te$_3$ \cite{Coelho13}. The fraction of the surface covered by islands increased until stabilizing around 10\% after 2 hours of air exposure when island formation ceased due to oxidation of \BS.

As a surface-sensitive probe, angle-resolved photoemission spectroscopy (ARPES) is well suited to characterize the surface electronic structure. However, high quality ARPES data of Bi surfaces on air-exposed \BS{} has not been reported yet. Rather, previous ARPES measurements of \BS{} exposed to air for 10~s showed a topological surface state similar to vacuum-cleaved \BS{} but with significant n-doping \cite{Analytis10}. For longer exposures of less than five minutes, quantum well states have been observed that coexist with the topological surface state \cite{Benia11,Chen12}.

Here, we present ARPES data of \BS{} exposed to air for 1 hour, long enough to allow for the formation of Bi surfaces with their own distinct band structures. These results clarify the air exposure time necessary for Bi surface formation, and allow for comparison to calculated band structures of Bi-surfaced \BS{} from literature \cite{Eich14}.

\section{Experiment}

Single crystals of \BS{} were grown by slow directional solidification from a binary melt. Stoichiometric quantities of the elements in the finest available powder form and of $99.999$\% or higher purity were loaded in a quartz tube and sealed under low pressure argon gas. The quartz ampoule was placed in a 1500~$^\circ$C Lindberg box furnace and deliberately off-centered such that it rests in a slight temperature gradient that was expected to be $<0.5-1$~K/cm. The tube was then heated to 750~$^{\circ}$C and after 12 hours was cooled to 650~$^{\circ}$C at a rate of 3~K/hour. This method combines aspects of slow cooling and Bridgman growth. To ensure reproducibility, samples from two different growth batches were used. Samples A and B are from one batch, samples C and D are from another.

\BS{} samples were cleaved in atmosphere unless indicated otherwise. Samples were mounted with Epotek H20E epoxy and cleaved by gluing a ceramic post to the surface, then knocking off the post. Typical relative humidity levels in our laboratory range from $40-60$\%. After 1 hour of air exposure samples were inserted into the load-lock of our vacuum system and pumped down to  $<5\times10^{-6}$~Torr within 5 minutes and $<5\times10^{-8}$~Torr within 24 hours. To avoid annealing effects at elevated temperatures \cite{Coelho13,Schouteden16,Gonalves18}, no bake-out was performed and the samples were retained at room temperature until and throughout measurement. 

$6$~eV photons were generated from a Ti:Sa oscillator operating at a repetition rate of $80$~MHz and two stages of second harmonic generation in $\beta$-BaB$_2$O$_4$ nonlinear crystals \cite{Kato86,Gauthier2020}. The diameter of the beam on the sample was $25$~$\mu$m FWHM. Photoelectrons were collected with a hemispherical analyzer.We estimate an instrumental energy resolution of $8$~meV. Unless otherwise specified, samples were biased at up to $-60$~V, increasing the momentum range of electrons collected by the analyzer \cite{Ichihashi18,Yamane19,Pfau20,Gauthier2021}. The light was p-polarized with the scattering plane perpendicular to the analyzer slit. ARPES measurements were performed at room temperature at pressures under $10^{-10}$~Torr. 

\section{Results}

In Figure~\ref{fig:arpes}(c) we plot a typical ARPES spectrum of vacuum-cleaved \BS. The spectrum of the vacuum-cleaved material is dominated by two topological surface state bands which intersect at a Dirac point $0.2$~eV below the Fermi level ($E_F$), with bulk states at higher binding energies. The position of the Dirac point below $E_F$ indicates n-type doping. This spectrum is consistent with calculated band structures for \BS{} \cite{Sobota13,Govaerts14} and numerous ARPES measurements \cite{Xia09,Analytis10,Sobota14}.

Air-cleaved samples display qualitatively different bands as shown in Figure~\ref{fig:arpes}(d) for a \BS{} surface that was exposed to ambient air for 1 hour. This data was taken during the same experimental run and on same crystal. Only the cleaving environment differed: Sample A1 was cleaved in vacuum and sample A2 was cleaved in air. The band structure is dominated by one inner and two outer bands  with hole-like dispersion centered at the $\Gamma$-point. These features are not as sharp as the topological surface state in a vacuum-cleaved specimen yet present well-defined bands worthwhile of a more detailed investigation. Energy distribution curves from the vacuum- and air-cleaved spectra in Figure~\ref{fig:arpes}(e) confirm that both surfaces present metallic states near $E_F$ that follow a Fermi-Dirac distribution for $300$~K. 

\begin{figure*}
\includegraphics[width=\textwidth]{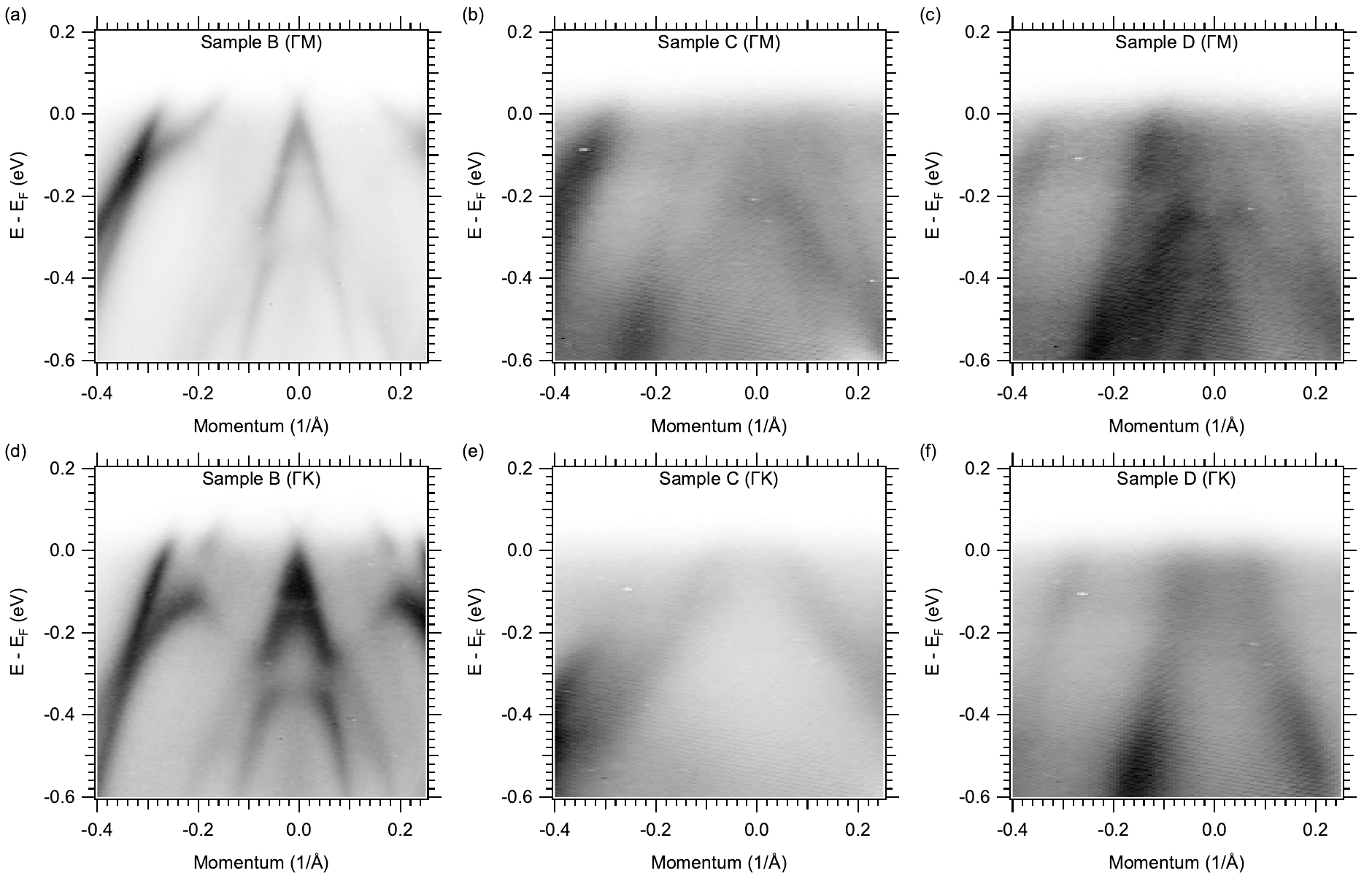}
\caption{ARPES data of \BS{} cleaved in and exposed to ambient air for 1 hour. This overview exemplifies variations between samples and high symmetry momentum directions. (a-c) in the $\Gamma$-M direction; (d-f) in the $\Gamma$-K direction. (a,d) are from Sample B; (b,e) are from Sample C; (c,f) are from Sample D.}
\label{fig:overview}
\end{figure*}

In Figure~\ref{fig:overview}, we present an overview of the variation between samples along two high-symmetry directions, $\Gamma$-K and $\Gamma$-M, see Figure~\ref{fig:overview}(b). These orientations refer to the orientation of the \BS{} sample as measured using Laue diffraction. There are significant differences in ARPES spectra between samples; this highlights how air-cleaving \BS{} is a stochastic process that results in quantitatively varying band structures. However, the qualitative differences between air- and vacuum-cleaved specimens remain obvious in all instances.

\begin{figure*}
\includegraphics[width=\textwidth]{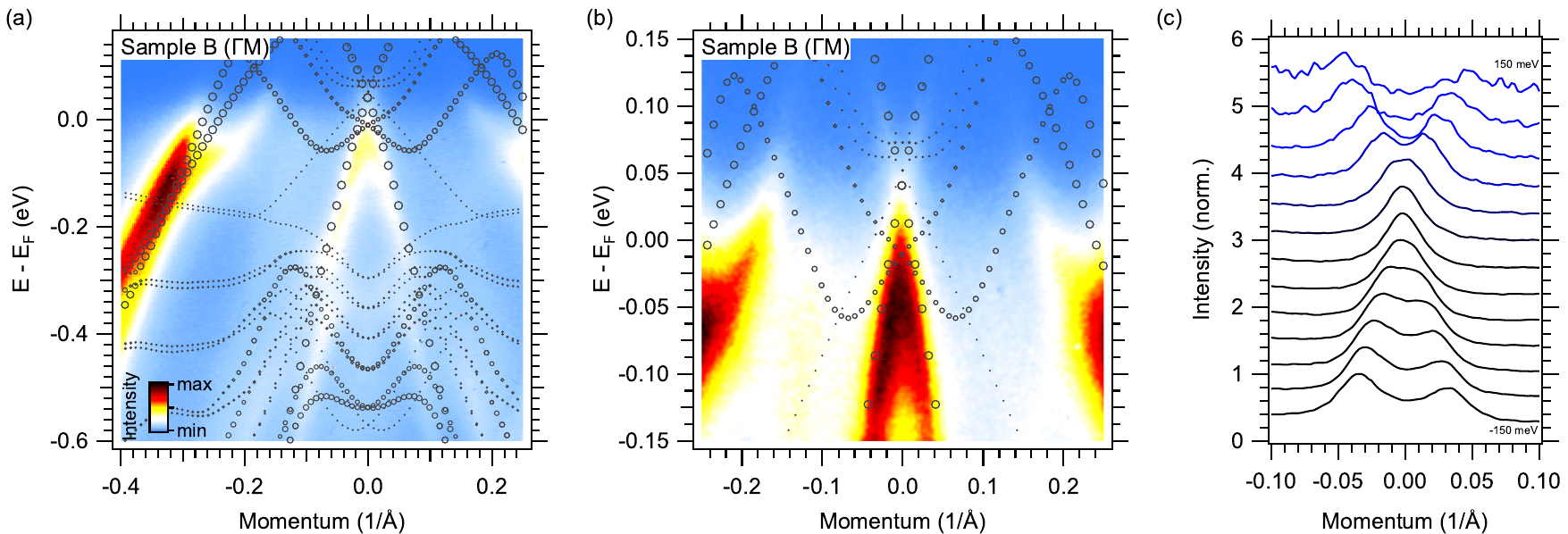}
\caption{(a) The same data as displayed in Figure~\ref{fig:overview}(a) is compared to calculated band structures of a single bilayer of Bi on \BS{} from Eich \textit{et al.} \cite{Eich14}. The theory data was shifted lower in binding energy by $0.25$~eV. (b) same as (a) zoomed in at $\Gamma$ and $E_F$. (c) Momentum distribution curves were taken from $-150$~meV to $150$~meV around $E_F$ with a $25$~meV integration window. The Dirac point of the topological surface state is located at $E_F-0.05$~eV.}
\label{fig:theory}
\end{figure*}

The data from Sample B in Figure~\ref{fig:overview}(a,d) is dominated by two outer hole bands and an inner, mostly linearly dispersing band. As we show in detail in Figure~\ref{fig:theory}, this inner band features a band crossing near $E_F$ and we assign this band to the topological surface state. A weak electron band hybridizes with the surface state near $E_F-0.3$~eV. A smaller decrease in intensity of the surface state near $E_F-0.15$~eV may be caused by a second hybridization.

Figure~\ref{fig:overview}(b,c,e,f) show data from Samples C and D, which are somewhat different from Samples A and B. A different doping shifts the Dirac point and hybridization features to above $E_F$. The two spectra in the $\Gamma$-M direction show signatures of an additional hole band at higher binding energies. In Figure~\ref{fig:overview}(c) this band displays splitting.

\section{Discussion}

We attribute the spectra observed in air-cleaved \BS{} to Bi surfaces on \BS{}, in accordance to previous observations with XPS and AFM \cite{Edmonds14,Hewitt14,Green16}. The simplest model to which we can compare our data is a single freestanding bilayer of Bi. Calculated band structures for this system \cite{Wang13,Hirahara11,Liu11,Govaerts14,Bieniek17}  are dominated by two hole bands centered at $\Gamma$ with similar velocity to our measured bands, providing good agreement with the major features of our measured data. However, this simple model is insufficient to explain our observation of a Dirac point and hybridization effects.

To explain these features we must turn to a slightly more complicated model. Bi-surfaced \BS{} \cite{He2013,Eich14,Lei16,Su17} and Bi$_2$Te$_3$ \cite{Hirahara11,Yang12,Miao13} have been extensively studied in the literature, allowing us to compare our spectra to the calculated band structure of Bi on \BS{}. Density functional theory calculations by Eich \textit{et al.} \cite{Eich14} of a single bilayer of Bi on \BS{} are overlaid on our data in Figure~\ref{fig:theory}(a,c). The calculated bands are shifted to lower binding energy by $0.25$~eV to match our measurements. This modest energy shift can be explained by differences in doping.

The calculations reproduce the most salient features in our data: two outer hole bands, an inner topological surface state, electron bands crossing the topological state slightly below the Dirac point, and an additional hole-like band at higher binding energy. The topological state is predicted to be strongly localized within the Bi surface layer. The outer hole bands are assigned to Bi valence bands, the electron bands crossing the topological state to \BS{} conduction bands, and the hole-like band at higher binding energy to \BS{} valence bands. Approximate band velocities are reproduced. While the calculations predict that the \BS{} conduction bands cross the topological state just below the Dirac point, they do not predict the hybridization we see in our data.

The calculations do not reproduce the split band we observe below $-0.3$~eV, see Figure~\ref{fig:overview}(c). This feature can be attributed to a substrate effect, where an electric field from the \BS{} substrate breaks the spin degeneracy at the surface. This leads to Rashba spin splitting of the bands in the Bi surface bilayer \cite{Wang13}.

Our spectra also share similarities with ARPES measurements of Bi$_2$-terminated Bi$_4$Se$_3$, which also feature topological surface states with a Dirac point near $E_F$, and an electron band crossing several hundred meV below the Dirac point \cite{Gibson13}.

\begin{figure}
    \centering
    \includegraphics[width=0.48\textwidth]{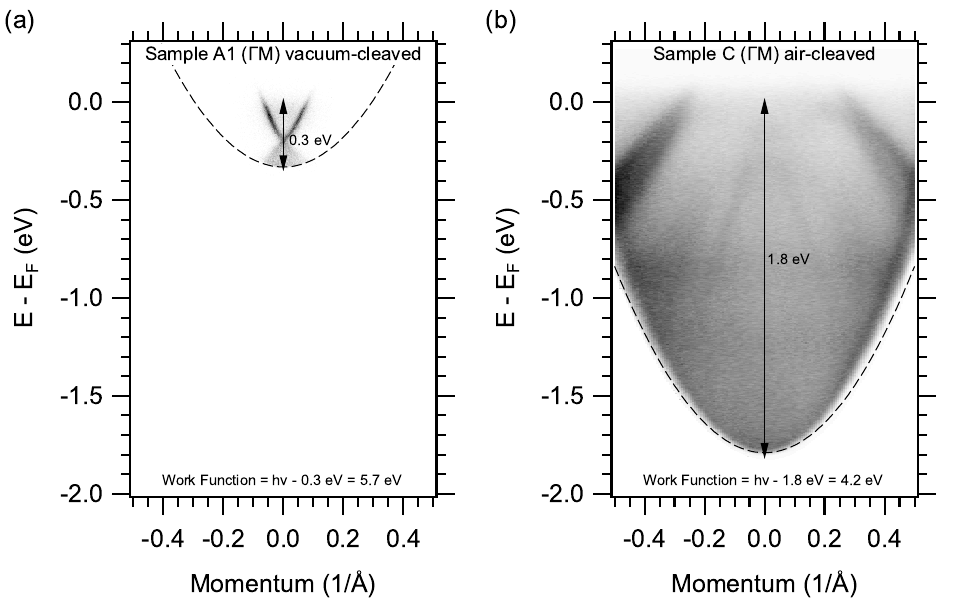}
    \caption{ARPES data covering the full emission range at $6$~eV photon energy is used to quantify the sample work function \cite{Pfau20}. The dashed lines indicate the kinematic limit of the photoemission cone for a given photon energy and sample work function. The vacuum-cleaved sample has a work function of $5.7$~eV while the air-cleaved specimen has a work function of $4.2$~eV.} 
    \label{fig:workFunc}
\end{figure}

Work function measurements further corroborate that the bands in air-cleaved \BS{} originate from a Bi dominated surface. In Figure~\ref{fig:workFunc} we show ARPES spectra with an extended energy range which allows determining the sample work function \cite{Pfau20}. The work function of this air-cleaved \BS{} sample is $4.2$~eV and similar to the work function of elemental Bi of $4.34$~eV \cite{CRC19}. In contrast, the vacuum-cleaved \BS{} sample has a considerably higher work function of $5.7$~eV. This is in good agreement with previous work \cite{Takane16,Sobota13}. As expected for the Se termination of \BS{} cleaves, this value is close to the work function of elemental Se of $5.9$~eV \cite{CRC19}.

Our ARPES measurements cannot determine the fractional Bi surface coverage. Our laser spot size was 25~$\mu$m; AFM results suggest that Bi islands on \BS{} can be on the order of 1~$\mu$m in size \cite{Green16}. Within the limits of our spatial resolution, every sample exposed to air for 1~hour displayed Bi bands over some portion of the sample. Some had Bi bands at only one or two spots, others had Bi bands covering the entire surface. Regions of the surface which displayed no Bi bands had negligible photoemission counts compared to the Bi regions and appeared passivated. Notably, band structures attributable solely to \BS{} were never observed on any surface exposed to air for 1~hour. This could be because the entire surface changed to Bi, or because remaining \BS{} areas were occluded by surface adsorbates or oxidation. Samples exposed to air for only 5 minutes did not display any Bi bands, in agreement with AFM and earlier ARPES measurements \cite{Benia11,Chen12,Green16}. Rather the usual but broadened \BS{} band structure was observed.

Bi bilayers at the surface are not expected to oxidize at room temperature \cite{Hapase67,Tahboub79,Ullmann00}. Depending on sample quality, Se-terminated regions of the sample may or may not oxidize after weeks to months in air \cite{Yashina13,Atuchin11,Thomas15,Kong11}. Oxidation timescales of \BS{} have been reported ranging from years to ten seconds \cite{Yashina13,Atuchin11,Thomas15,Kong11,Hong2020} depending on growth conditions and cleavage methods \cite{Golyashov12}.

\section{Conclusions}

We report that exposure of \BS{} to air for 1 hour consistently results in the formation of ordered Bi bilayers with well-defined electronic band structures that are readily observed with photoemission. These band structures are well explained by a single bilayer of Bi on \BS{}, and are observed to the exclusion of typical \BS{} spectra.

We stress that air-exposed \BS{} is not a clean system. Stochastic variations from sample to sample are reflected in varying photoemission  intensities, band velocities, and dopings. The photoemission signal also has a diffuse, momentum-independent background we attribute to scattering from air adsorbates and the formation of Bi islands. The presence of air adsorbates gives laser-based $6$~eV photoemission an advantage over measurements at higher photon energies, due to the longer electron mean free path and thus decreased surface sensitivity at $6$~eV \cite{Seah1979, Koralek07}. However, even at $6$~eV the high data quality of vacuum-cleaved samples cannot be attained.

We conclude that exposing \BS{} to air is not a well-defined way to generate structures of Bi on \BS{} for detailed studies, which requires controlled deposition of Bi on \BS. But it remains an interesting question how these air-induced Bi surface structures impact the surface and topological properties of \BS {} as we observe hybridization between \BS{} conduction states and the topological state. Since the observation of two-dimensional conduction channels in quantum oscillations is a hallmark of topological surface states \cite{Analytis2010,Li14}, it is particularly important to consider the influence of 2D transport channels coexisting with those from the topological surface states \cite{Bansal12}. These considerations are also relevant for surface-sensitive experiments performed in air, such as surface second harmonic generation \cite{Hsieh11_SHG} or high-harmonic generation \cite{Baykusheva21,Heide2022}. Additionally, few-layer Bi thin films are 2D topological insulators in their own right \cite{Liu11,Wang13}. It is also conceivable that transport experiments and device applications may be impacted by the presence of Bi metal at the surface when contacts and devices are manufactured in air. To avoid the complications caused by Bi surface layers, surface-sensitive measurements of \BS{} should be performed on fresh surfaces which have not been exposed to air for more than a few minutes. Practitioners should keep in mind that Bi surfaces form on \BS{} upon exposure to air of tens of minutes.

\begin{acknowledgements}
    This work was supported by the U.S. Department of Energy (DOE), Office of Science, Basic Energy Sciences, Materials Sciences and Engineering Division under Contract No. DE-AC02-76SF00515. A.G. acknowledges support from the Stanford Graduate Fellowship.
\end{acknowledgements}

\end{document}